\documentclass[conference]{IEEEtran}
\IEEEoverridecommandlockouts

\usepackage{cite}
\usepackage{amsmath,amssymb,amsfonts}
\usepackage{algorithm}
\usepackage{algpseudocode}
\usepackage{listings}
\usepackage{graphicx}
\graphicspath{{images/}}
\usepackage{textcomp}
\usepackage{xcolor}
\usepackage{subfig}
\def\BibTeX{{\rm B\kern-.05em{\sc i\kern-.025em b}\kern-.08em
    T\kern-.1667em\lower.7ex\hbox{E}\kern-.125emX}}

\begin{document}

\title{Interference Mitigation Scheme in 3D Topology IoT Network with Antenna Radiation Pattern
\thanks{This work is supported in part by the INL Laboratory Directed Research Development (LDRD) Program under DOE Idaho Operations Office Contract DEAC07-05ID14517, and by NSF CNS through award number 1814727.}}

\author{
\IEEEauthorblockN{Sung Joon Maeng$^*$, Mrugen A. Deshmukh$^*$, \.{I}smail G\"{u}ven\c{c}$^*$, and  Arupjyoti Bhuyan$^\dagger$}
\IEEEauthorblockA{$^*$Department of Electrical and Computer Engineering, North Carolina State University, Raleigh, NC\\
$^\dagger$Idaho National Laboratory, Idaho Falls, ID\\
\{smaeng, madeshmu, iguvenc\}@ncsu.edu, arupjyoti.bhuyan@inl.gov}}

\maketitle

\begin{abstract}
Internet of things (IoT) is one of main paradigms for 5G wireless systems. Due to high connection density, interference from other sources is a key problem in IoT networks. Especially, it is more difficult to find a solution to manage interference in uncoordinated networks than coordinated system. In this work, we consider 3D topology of uncoordinated IoT network and propose interference mitigation scheme with respect to 3D antenna radiation pattern. In 2D topology network, the radiation pattern of dipole antenna can be assumed as onmi-directional. We show the variance of antenna gain on dipole antenna in 3D topology, consider the simultaneous use of three orthogonal dipole antennas, and compare the system performance depending on different antenna configurations. Our simulation results show that proper altitude of IoT devices can extensively improve the system performance.  
\end{abstract}


\section{Introduction} \label{S1}
The growing demand on Internet of things (IoT) devices for
next-generation wireless systems requires massive number of connections in 5G communications~\cite{1},\cite{2}. Heterogeneous network is considered on IoT network in order to satisfy pervasive, always-on, dense connectivity \cite{3,merwaday2016handover,merwaday2014capacity}. In such dense uncoordinated network, interference management problem is significant. Besides, interference mitigation technique should be feasible and low-cost for distributed IoT networks.\par

Depending on applications and scenarios, realistic 3D topology model can be appropriate in IoT networks \cite{5}, \cite{6}. For example, the use of unmanned aerial vehicles (UAVs) combined with IoT devices is considered in the future wireless network \cite{7}. In our paper, we consider 3D topology IoT network and assume that IoT devices are located at high altitude position. In this scenario, we propose interference mitigation strategy by considering the impact of 3D antenna radiation pattern of dipole antenna with different configurations. The 3D radiation pattern of single dipole antenna~\cite{chen2018impact} and two cross-dipole antenna~\cite{8,9} configuration are  studied in the literature. Unlike 2D domain geometric setup, the radiation pattern of dipole antenna varies depending on the location of devices. We study the feasibility of single dipole antenna in the 3D topology and propose new strategy for antenna configuration. We show that we can utilize the directivity of antenna radiation pattern to suppress received signal power to other IoT devices.\par

\begin{figure}[t]
     \centering
     	\subfloat{
     	\includegraphics[width=0.9\linewidth]{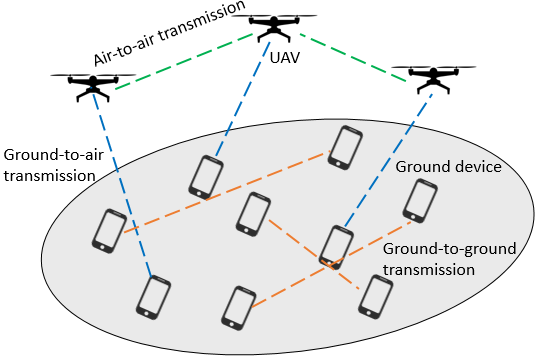}}
        \caption{Illustration of 3D topology uncoordinated IoT network.}
			\label{ill} \vspace{-4mm}
\end{figure}

A major use case of IoT devices is sensor networks and environmental data collection \cite{10}. For this reason, low power consumption and small size of devices are needed. We consider the number of dipole antennas from 1 to 3 with low transmission power. In addition, in order to take advantage of the directivity of antenna radiation pattern, transmitters need to decide which dipole antennas should be used to increase the link quality. We propose antenna selection methods based on detecting and power measurement of preambles on the side of transmitters. Based on our simulation results, it is observed that 2 or 3 dipole antenna configuration can achieve higher overall capacity of the network compared with single dipole configuration by the benefit of the diversity of antenna radiation pattern. In addition, we observed that higher altitude position of devices can contribute to better performance when the devices are uniformly distributed on ground and air.

\section{System Model}  \label{S2}
We consider an uncoordinated IoT network where point-to-point (p2p) transmitter and receiver IoT device pairs are distributed in an area as in Fig.~\ref{ill}. As lliustrated in Fig.~\ref{ill}, there are three kind of links; air-to-air, ground-to-air, ground-to-ground. In simulation results section \ref{simulation}, it is considered that all links are reflected in separate scenarios. Since all p2p pairs are not scheduled by base station, they share the same time and frequency resources in the network. 

\subsection{3D Topology Based Channel Model}
We design channel model based on 3D coordinate position of transmitter and receiver devices. The distance between i-transmitter and i-receiver is defined by $d_{i,i}=\sqrt{(x^{\rm{Tx}}_i-x^{\rm{Rx}}_i)^2+(y^{\rm{Tx}}_i-y^{\rm{Rx}}_i)^2+(z^{\rm{Tx}}_i-z^{\rm{Rx}}_i)^2}$. Then, free-space pathloss can be expressed as
\begin{align}\label{1}
\beta_{i,i}&= \left(\frac{\lambda}{4\pi d_{i,i}}\right)^2,
\end{align}
where $x_i, y_i,z_i$ is 3D coordinate position of i-device, and $\lambda$ is the wave length of the signal. P2p channel between i-transmitter and i-receiver can be modeled as
\begin{align}\label{2}
h_{i,i}&= \sqrt{P_{\rm{Tx},i}G_i^{\rm{Tx}}\beta_{i,i}G_i^{\rm{Rx}}}\alpha_{i,i},
\end{align}
where $P_{\rm{Tx},i}$ is transmitted signal power of i-transmitter, $G_i^{\rm{Tx}}$, $G_i^{\rm{Rx}}$ are the antenna gain of i-transmitter and i-receiver, and $\alpha_{i,i}\sim \mathcal{CN}(0,1)$ is fading path gain. We assume single LoS path model. Note that channel gain varies depending on the relative location between Tx and Rx. Path-loss ($\beta_{i,i}$) is calculated by distance between Tx and Rx, and Tx, Rx antenna gain $G_i^{\rm{Tx}}, G_i^{\rm{Rx}}$ are changed by the angle of departure and arrival of signal.
\subsection{Antenna Radiation Pattern of Dipole Antenna}
\begin{figure}[h]
     \centering
     	\subfloat[z-axis dipole.]{
			\includegraphics[width=0.3\linewidth]{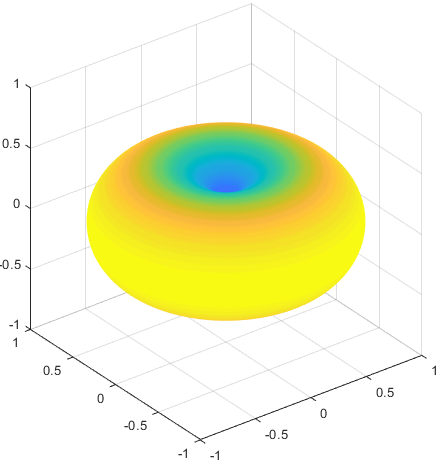}}
        \subfloat[y-axis dipole.]{
			\includegraphics[width=0.3\linewidth]{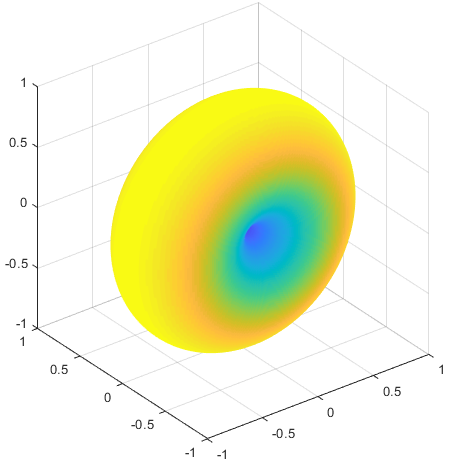}}
		\subfloat[y,z-axis dipoles.]{
			\includegraphics[width=0.3\linewidth]{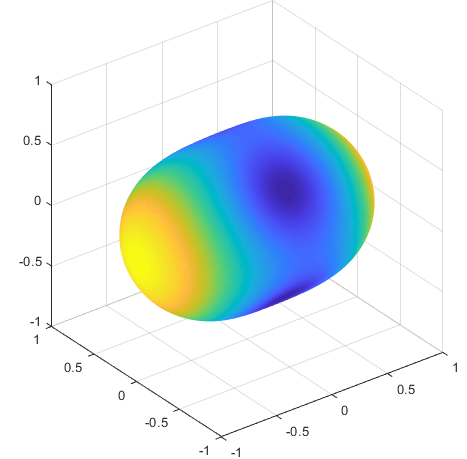}}
			
		\subfloat[x,z-axis dipoles.]{
			\includegraphics[width=0.3\linewidth]{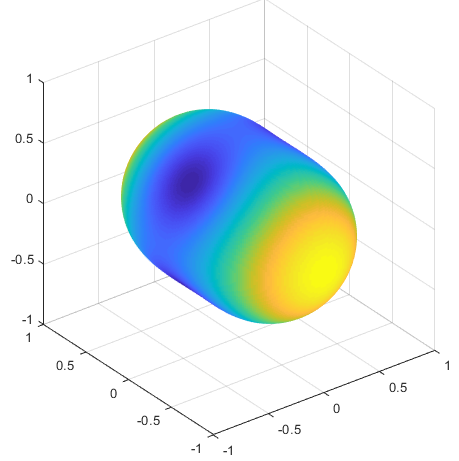}}
		\subfloat[x,y-axis dipoles.]{
			\includegraphics[width=0.3\linewidth]{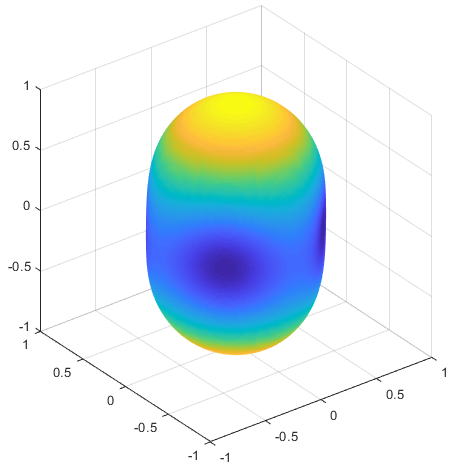}}
			\caption{Radiation pattern of various dipole antennas settings.}
			\label{pattern}
 \end{figure}

Antenna radiation pattern is different depending on antenna type. Dipole antenna has omni-directional radiation pattern in azimuth angle domain. Thus, we can easily assume constant antenna gain in 2D topology. However, in 3D topology scenario, the radiation pattern of dipole antenna varies depending on elevation angle. The radiation pattern of z-axis dipole antenna can be written as
\begin{align}\label{3}
G_z(\theta,f_0)&=\frac{\cos\left(\frac{\pi d_{\rm{len}}}{c}f_0\cos\theta\right)-\cos\left(\frac{\pi f_0d_{\rm{len}}}{c}\right)}{\sin\theta},
\end{align}
where $\theta$ is elevation angle of departure, $f_0$ is carrier frequency, $d_{\rm{len}}$ is the length of dipole antenna, and $c\approx3\times10^8 m/s$ is the speed of light \cite{11}. Fig.~\ref{pattern}(a) shows the 3D radiation pattern of z-axis dipole antenna. It can be observed that antenna gain become low, as receiver is located at high altitude. We can also consider y-axis dipole configuration. It means that the dipole antenna is implemented to the direction of y-axis. In this case, the radiation pattern varies depending on both azimuth and elevation angle. The radiation pattern of y-axis dipole can be expressed as
\begin{multline}\label{4}
G_y(\phi,\theta,f_0)\\
=\frac{\cos\left(\frac{\pi d_{\rm{\rm{len}}}}{c}f_0\cos(\cos^{-1}(\sin\theta\sin\phi))\right)-\cos\left(\frac{\pi f_0d_{\rm{len}}}{c}\right)}{\sin(\cos^{-1}(\sin\theta\sin\phi))},
\end{multline}
where $\phi$ is azimuth angle of departure. The 3D radiation pattern of y-axis dipoles antenna is shown in Fig.~\ref{pattern}(b). We can see that antenna gain is high for near origin with high altitude IoT device, while a device that is located near x-axis with low altitude obtains low gain.\\
If we utilize two dipole antenna, we can achieve combined radiation pattern of antenna. Considering two cross-dipole antenna with x,y-axis dipole, The radiation pattern can be written as
\begin{multline*}
G_{xy}(\phi,\theta,f_0)\\
=\frac{1}{\sqrt{2}}\frac{\cos\left(\frac{\pi d_{\rm{len}}}{c}f_0\cos(\cos^{-1}(\sin\theta\cos\phi))\right)-\cos\left(\frac{\pi f_0d_{\rm{len}}}{c}\right)}{\sin(\cos^{-1}(\sin\theta\cos\phi))}\\
+\frac{1}{\sqrt{2}}j\frac{\cos\left(\frac{\pi d_{\rm{len}}}{c}f_0\cos(\cos^{-1}(\sin\theta\sin\phi))\right)-\cos\left(\frac{\pi f_0d_{\rm{len}}}{c}\right)}{\sin(\cos^{-1}(\sin\theta\sin\phi))}~.
\end{multline*}
The radiation pattern of x,y-axis dipole is calculated by the sum of x-axis and y-axis dipole antenna gain with transmitted power normalization to 1. In a similar way, the radiation pattern of y,z-axis dipoles and x,z-axis dipoles can be derived easily. The 3D radiation pattern of y,z-axis, x,z-axis, x,y-axis dipoles is shown in Fig.~\ref{pattern}(c), (d), (e). The radiation pattern of two cross-dipole antennas have directivity toward certain direction depending on configuration.
\section{Interference mitigation scheme based on dipole antenna settings.}\label{S3}
In this section, we propose interference mitigation scheme based on antenna configuration. If we increase the number of antenna from 1 to 3, we can suppress interference to other devices by different radiation pattern. We divide schemes according to the number of dipole antennas.
\begin{figure}[h]
     \centering
     	\subfloat[Single dipole.]{
			\includegraphics[width=0.3\linewidth]{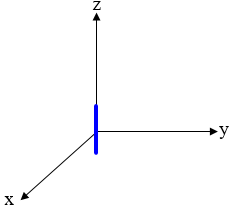}}
        \subfloat[2 orthogonal dipoles.]{
			\includegraphics[width=0.3\linewidth]{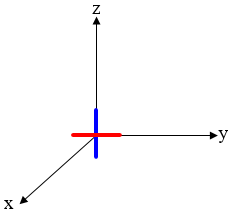}}\;
		\subfloat[3 orthogonal dipoles.]{
			\includegraphics[width=0.3\linewidth]{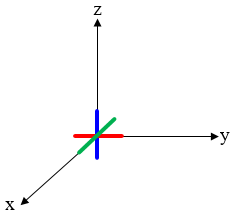}}
			\caption{Dipole antennas configuration.}
			\label{dipole}\vspace{-4mm}
 \end{figure}
\subsection{Single Dipole Antenna Scenario}
The dipole antenna placed along the z-axis is described in Fig.~\ref{dipole}(a). Single dipole antenna has omni-directional radiation pattern in the azimuth angle direction, while antenna gain decreases as elevation angle become smaller (Fig.~\ref{pattern}(a)). In 2D topology scenario, elevation angle can be neglected. Thus, omni-directional radiation pattern can be assumed. However, if we think of connection between ground transmitter with receiver in the air, transmitter requires more power in order to compensate the reduction of antenna gain. Besides, increasing transmitted power leads to more interference to nearby devices.

\subsection{Two Orthogonal Dipole Antennas Scenario}
We can consider devices with two orthogonal dipole antennas along with y, z-axis as shown in Fig.~\ref{dipole}(b). Since the devices have two antennas, they can choose either of two antennas depending on desired receiver's position. In other words, the radiation pattern of y-axis dipole antenna or the radiation pattern of z-axis dipole antenna radiation pattern (shown in Fig.~\ref{pattern}(a), (b)) can be selected by transmitters. Since z-axis dipole radiation pattern is suitable for ground devices, while y-axis dipole radiation pattern achieve higher antenna gain for high altitude devices, we can obtain higher overall system performance than 'single dipole scheme' in 3D topology network. The algorithm of how to select the best antenna at the transmitter is discussed in \ref{S3_4}.

\subsection{Three Orthogonal Dipole Antennas Scenario}
\begin{figure}[h]
     \centering
      	\subfloat[Single dipole devices.]{
			\includegraphics[width=0.45\linewidth]{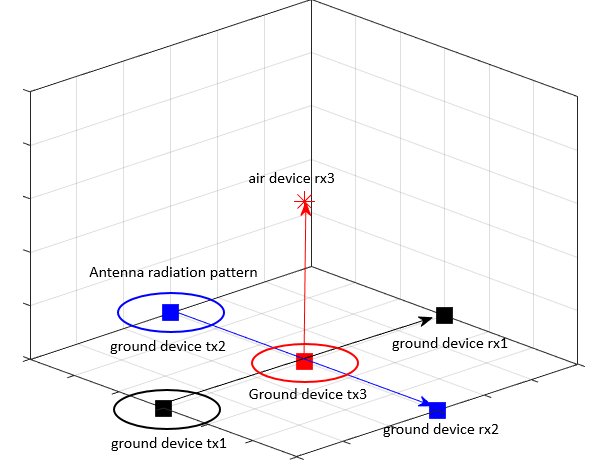}}
        \subfloat[3 orthogonal dipoles devices.]{
			\includegraphics[width=0.45\linewidth]{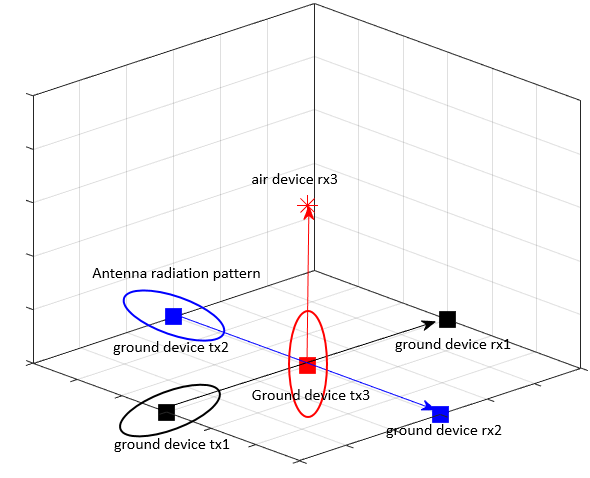}}
			\caption{An example of 3D topology scenario.}
			\label{deployment}\vspace{-4mm}
 \end{figure}

If we consider devices with three orthogonal dipole antennas, we can shape additional directional radiation patterns by utilizing two transmitted antennas out of three dipole antennas (the additional antenna radiation patterns are shown in  Fig.~\ref{pattern} (c), (d), (e)). Three orthogonal antennas placement is shown in Fig.~\ref{dipole}(c). Devices with three dipole antennas can choose the best radiation pattern which improves link qualities of network. Note that directional antenna radiation pattern can improve system performance by reducing interference power to other devices.
Fig.~\ref{deployment} shows an example of 3D topology communication scenario. In Fig.~\ref{deployment}(a), Every Tx device has single z-axis dipole antenna pattern. In Fig.~\ref{deployment}(b), Tx device selects the best directional radiation pattern to reduce interference toward other devices.
\subsection{Antenna Selection}\label{S3_4}
In this subsection, we propose antenna selection methods in order to find the best antenna pattern. In two, three dipole antenna scenario, transmitters need to decide which antennas they use. We consider that transmitters measure received signal power from uplink preambles. Let's assume that individual receivers transmit orthogonal preambles at the same time. If transmitters know transmitted preamble from the connected receivers, they can measure received signal power by cross-correlation operation (preamble signals from other receivers are eliminated by orthogonal property). In addition, if receivers transmit preambles repeatedly, and transmitter measures signal power by different antenna selections (change antenna selection per preamble period), the transmitter is able to decide the best antenna pattern by selecting the radiation pattern with the highest measured power of preamble. Let $\mathcal{G}$ be the candidate antenna selection, cardinality  $|\mathcal{G}|=2$ for two dipole antenna scenario, and $|\mathcal{G}|=6$ for three dipole antenna scenario (x, y, z, xy, yz, xz-axis antenna cases). The decision made by the highest received signal power (method 1) can be written as
\begin{align}\label{6}
    \{G^{\rm{Tx}}_{i}\}^{\boldsymbol{\star}}&=\arg \max_{\forall G^{\rm{Tx}}_{i}\in\mathcal{G}}P_{\rm{Tx},i}G_i^{\rm{Tx}}\beta_{i,i}\alpha^2_{i,i}~.
\end{align}

Another decision method is based on the signal-to-leakage-plus-noise-ratio (SLNR). SLNR measures energy leakage into undesired users as well as desired signal power. Similarly, we can consider SINR (Signal-to-interference-plus-noise-ratio) as a decision criteria. However, the advantage of SLNR measure is from the simplicity of measurement procedure. In order to measure SINR, individual receivers measure signal power from all transmitters and require to feedback measured information to the transmitter. Besides, interference can be changed depending on other transmitters antenna pattern decision, which makes difficult to decide antenna pattern without information of other transmitters' decision. On the other hands, SLNR can be measured at the transmitter's side, regardless of other transmitters' decision. The explanation of how to measure SLNR is as follow. All receivers send distinct orthogonal preambles repeatedly. If the transmitter knows all received preambles' sequence, it can measure separate preamble power. By cross-correlation of individual preambles, the transmitter can estimate desired signal power and leakage power to other receivers. The transmitter switches antenna selection per preamble, and measure preambles power. The decision method by SLNR can be expressed as
\begin{align}\label{7}
    \{G^{\rm{Tx}}_{i}\}^{\boldsymbol{\star}}&=\arg \max_{\forall G^{\rm{Tx}}_{i}\in\mathcal{G}}\frac{|h_{i,i}|^2}{\sum_{j\neq i}|h_{i,j}|^2+\sigma_n^2}~,
\end{align}
where $\sigma_n^2$ is the noise variance.\\

\section{Simulation Results} \label{simulation}

\begin{table}[h]
\centering
\caption{Simulation Parameters}\label{t1} 
\begin{tabular}{|l|c|} \hline
\textbf{Parameter} & \textbf{Value} \\ \hline\hline
Number of Tx and Rx pair ($K$)& 4  \\ 
x-coordinate of device & Uniform random in [-100, 100] m \\ 
y-coordinate of device & Uniform random in [-100, 100] m \\
z-coordinate of ground device & 0 m \\
z-coordinate of air device (height) & 150 m \\
Transmit power & 23 dBm \\
Carrier frequency($f_0$) & 800 MHz \\
Bandwidth ($B$)& 200 kHz \\
Noise power & $-174 + 10\log_{10}(B)$ dBm\\
Antenna gain of Rx ($G^{\rm{Rx}}$) & 1 (onmi-directional)\\
\hline
Channel state information & Ideal\\ 
Antenna selection mode & Method 1: Signal power based\\ &Method 2: SLNR based\\ \hline
\end{tabular}
\end{table}

\begin{figure}[]
     \centering
			\includegraphics[width=1\linewidth]{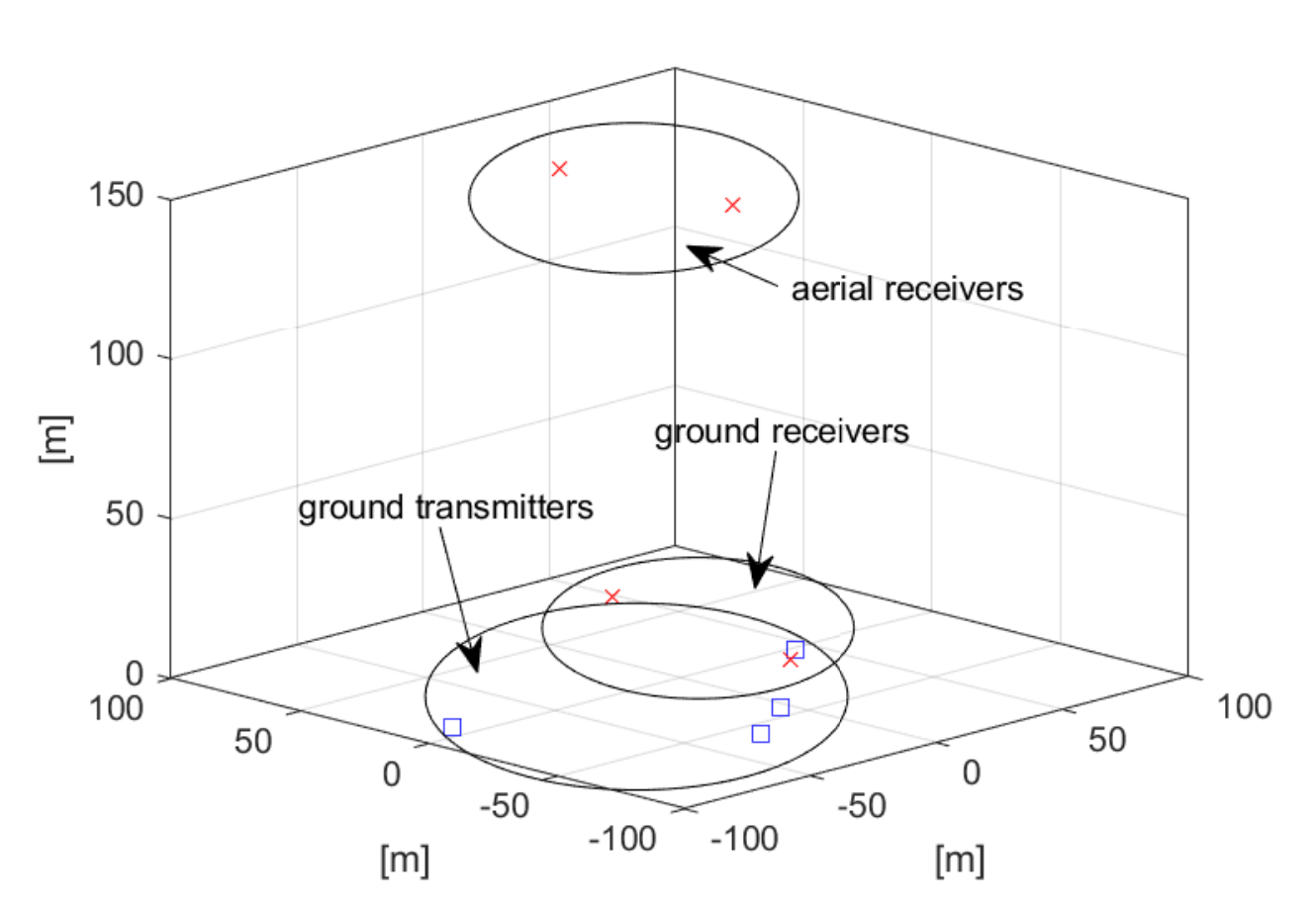}
        \caption{An experiment of 3D topology of IoT devices (height = 150 m, $K = 4$, air device percentage 50\%).}
			\label{posi}
 \end{figure}
 
In this section, we compare the performance of the proposed schemes by simulation. The simulation parameters are listed in the Table.~\ref{t1}. Fig.~\ref{posi} shows an experiment of 3D topology of IoT devices in simulations. We assume that transmitters are all ground users and receivers are distributed at both ground and air.\par

\begin{figure}[]
     \centering
     	\includegraphics[width=1\linewidth]{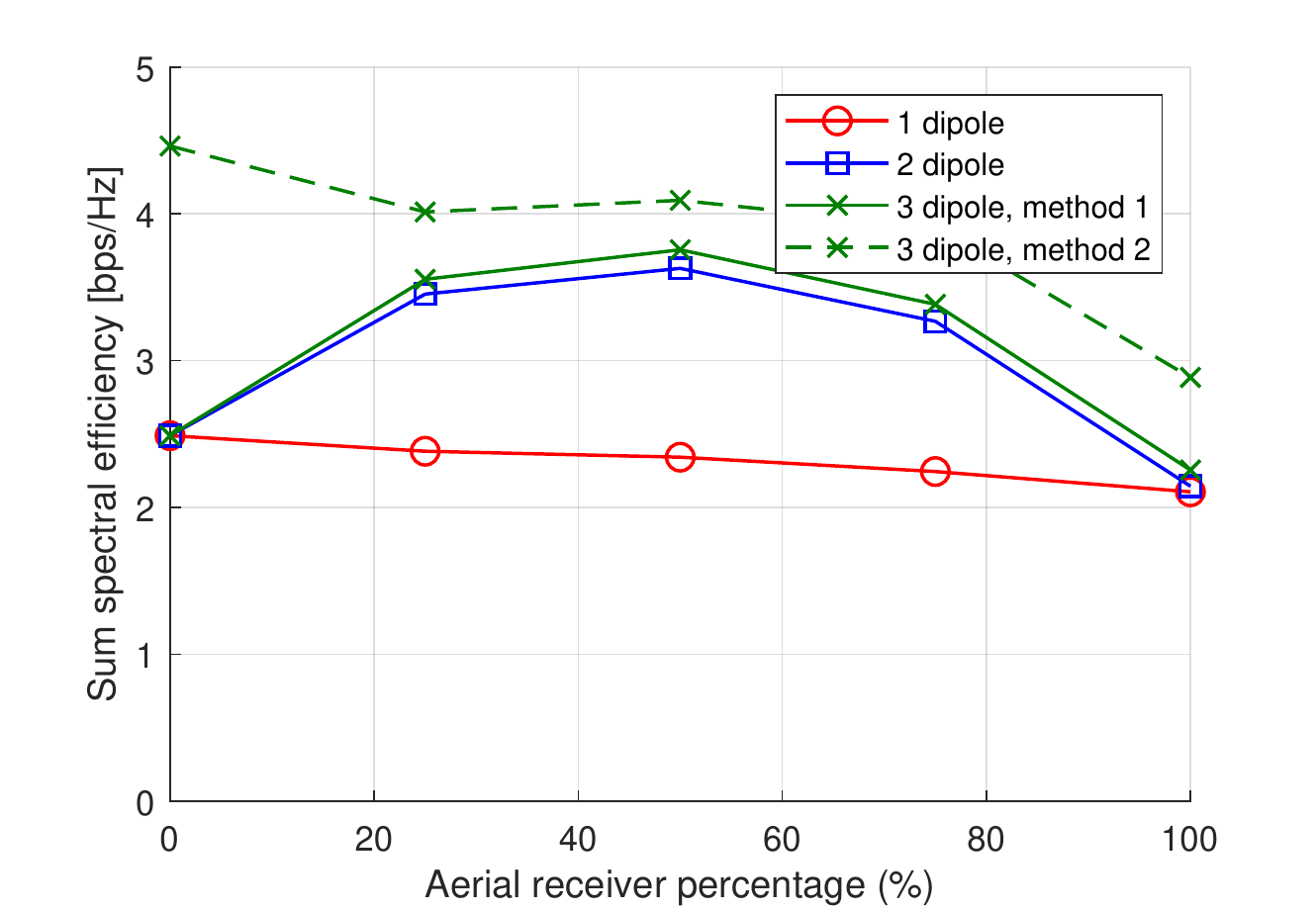}
        \caption{Sum spectral efficiency versus air device percentage.}
			\label{effi}
\end{figure}

In Fig.~\ref{effi}, we show sum achievable rate depending on ratio of the number of ground receivers to aerial receivers. For example, if aerial receiver percentage is 25, the number of aerial receivers is 1 out of 4 and ground receivers is 3 out of 4. The sum achievable rate is written as
\begin{align}\label{8}
   R_{\rm{sum}}&=\sum_{i=1}^K\log_2(1+\text{SINR}_i)\\
   &=\sum_{i=1}^K\log_2\left(1+\frac{|h_{i,i}|^2}{\sum_{j\neq i}|h_{j,i}|^2+\sigma_n^2}\right)~.
\end{align}
 \begin{figure}[]
     \centering
     	\subfloat[Desired signal power.]{
			\includegraphics[width=0.97\linewidth]{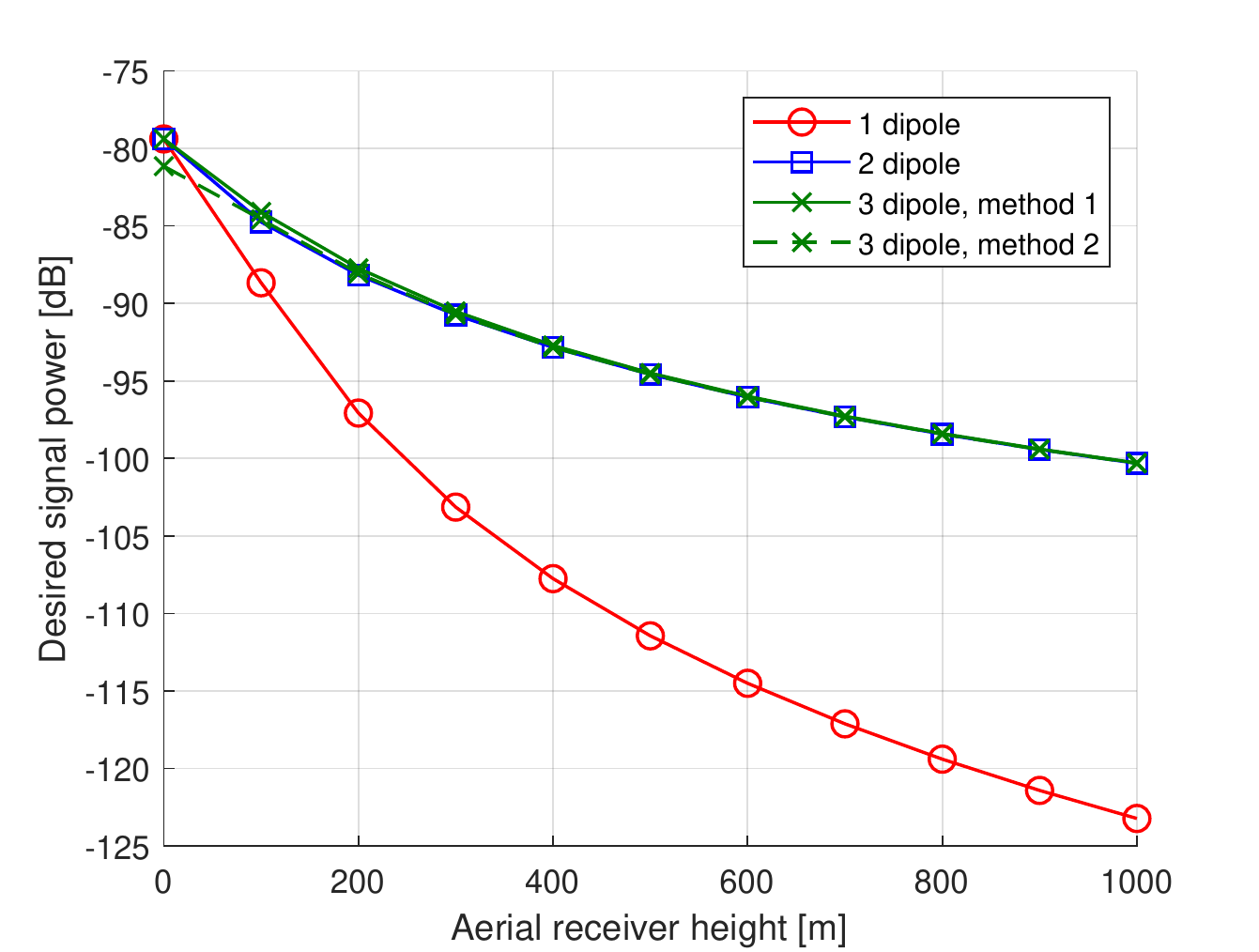}}
			
        \subfloat[Interference signal power.]{
			\includegraphics[width=0.97\linewidth]{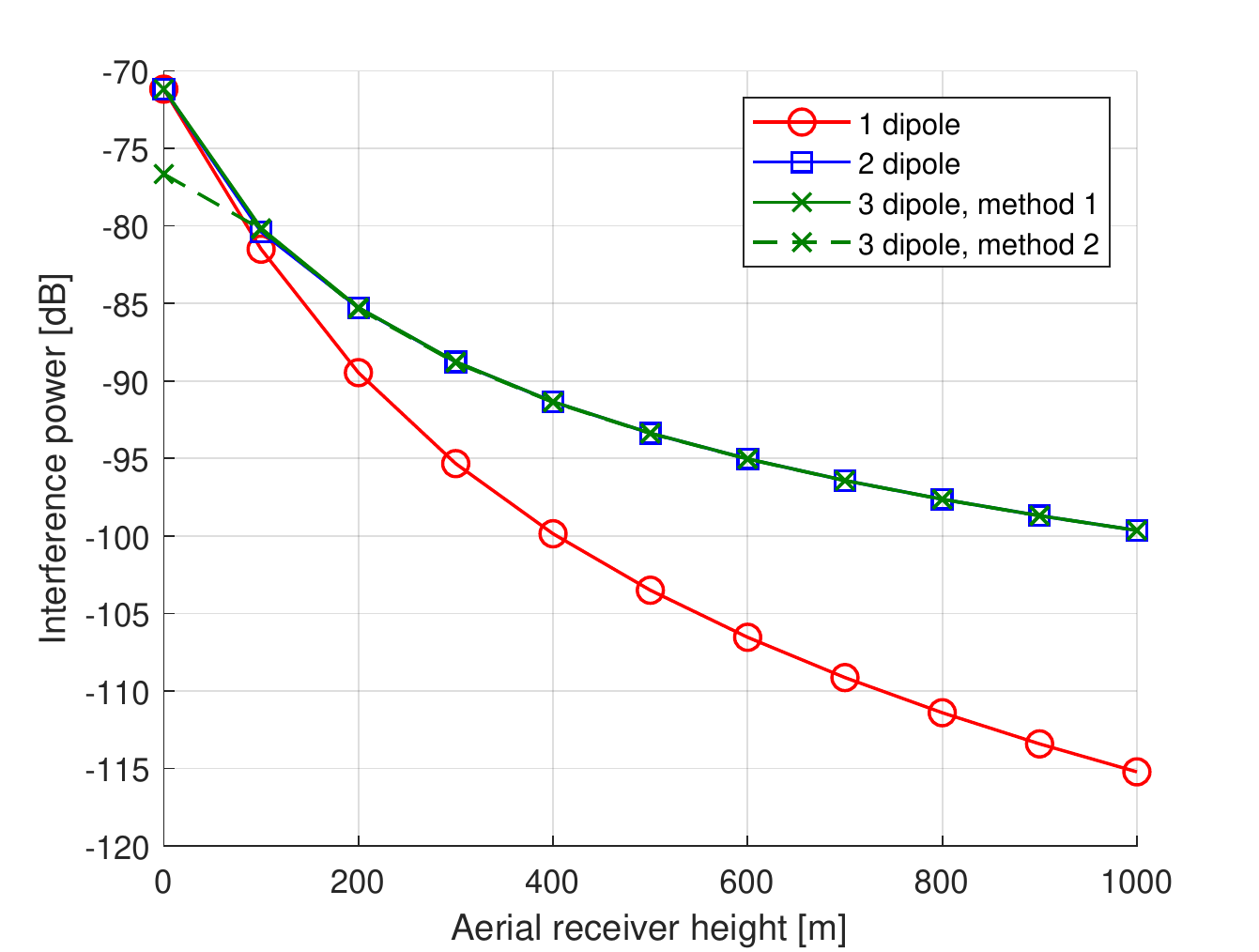}}
			
	        \subfloat[Signal-to-interference-ratio (SIR).]{
			\includegraphics[width=0.97\linewidth]{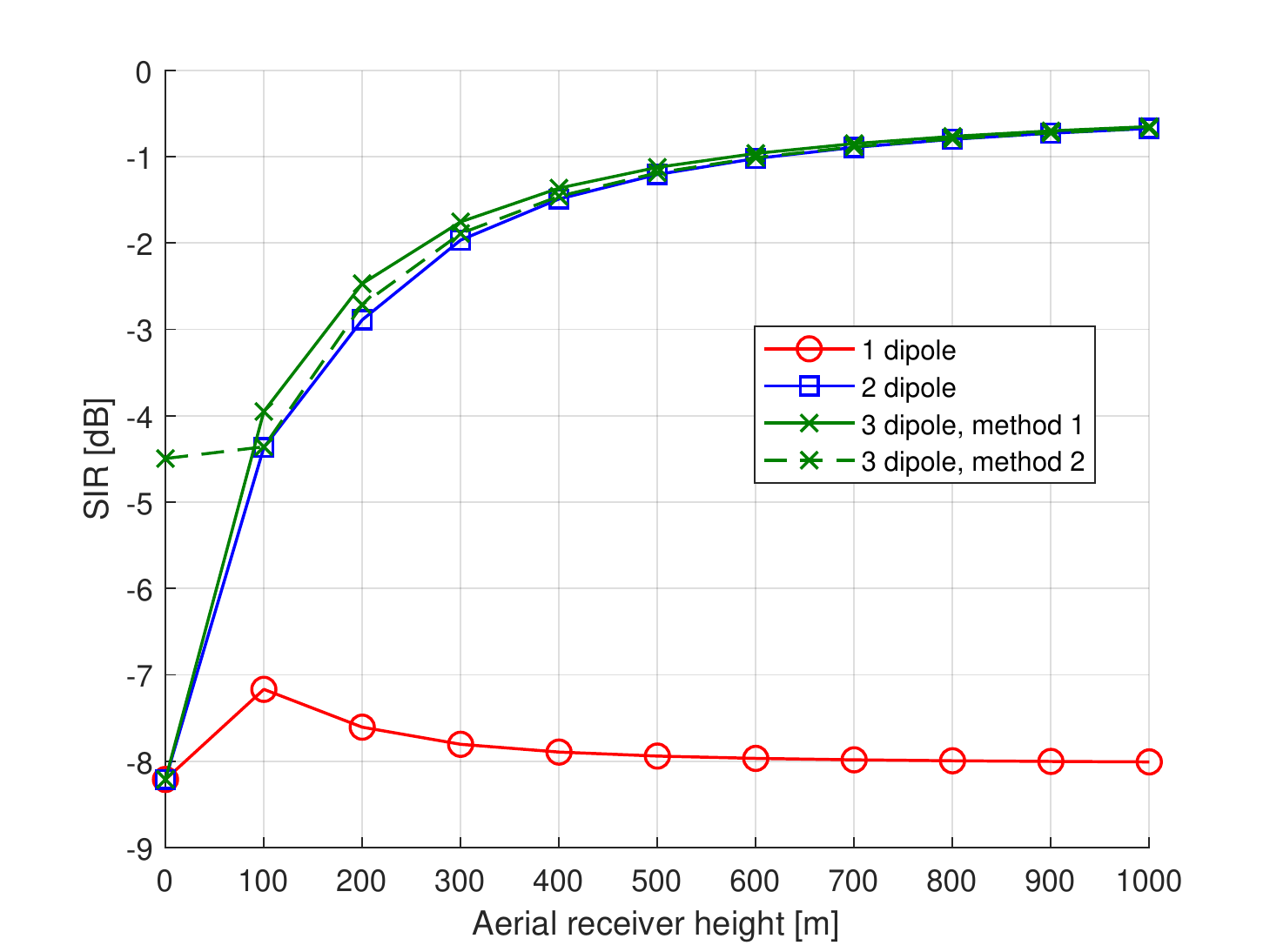}}
        \caption{Signal power of aerial receiver versus aerial receiver height ($K = 4$, air device percentage 50\%).}
			\label{pw}
\end{figure}
 \begin{figure}[t]
     \centering
     	\subfloat[Aerial receivers percentage: 50\%.]{
			\includegraphics[width=1\linewidth]{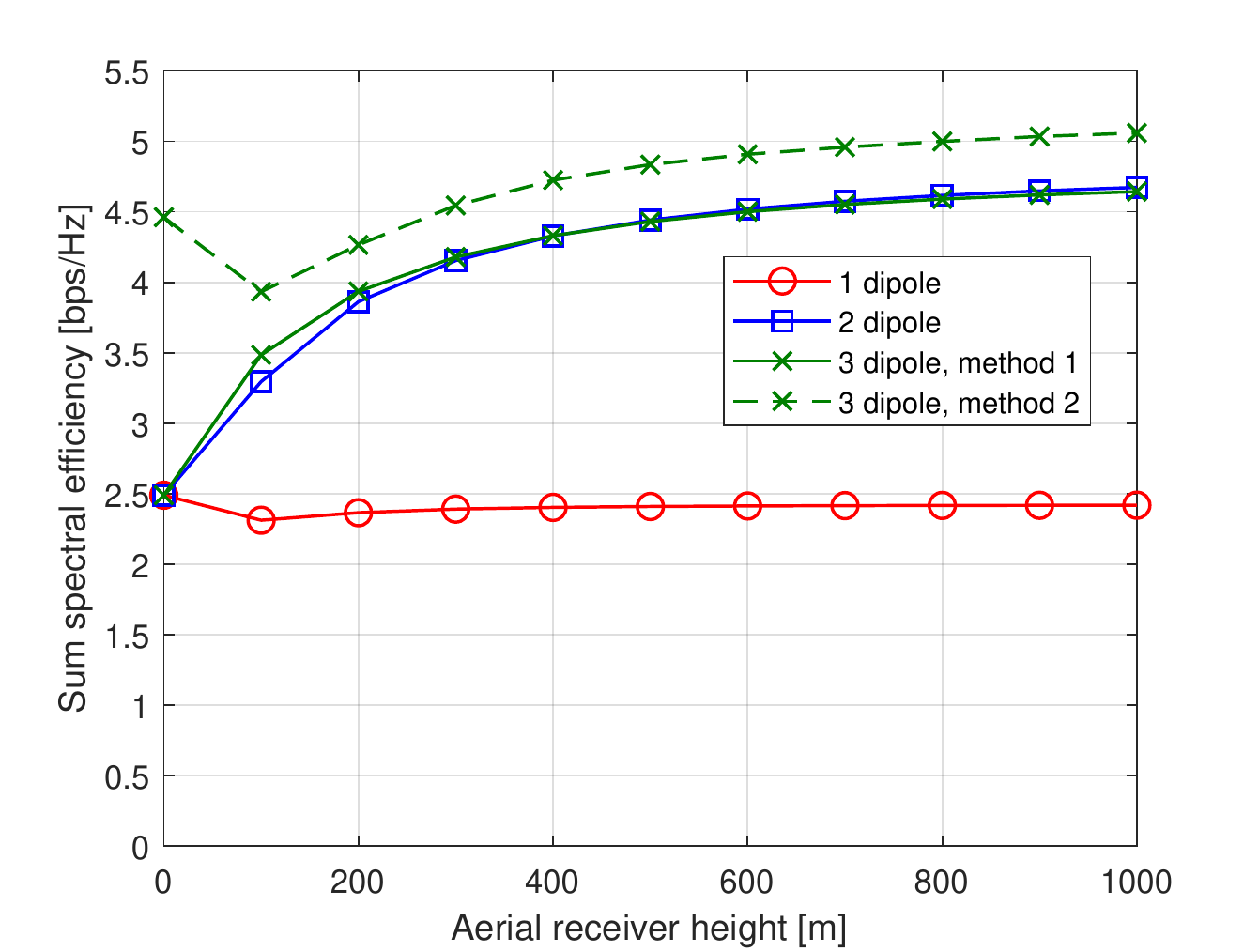}}
			
		\subfloat[Aerial receivers percentage: 75\%.]{	\includegraphics[width=1\linewidth]{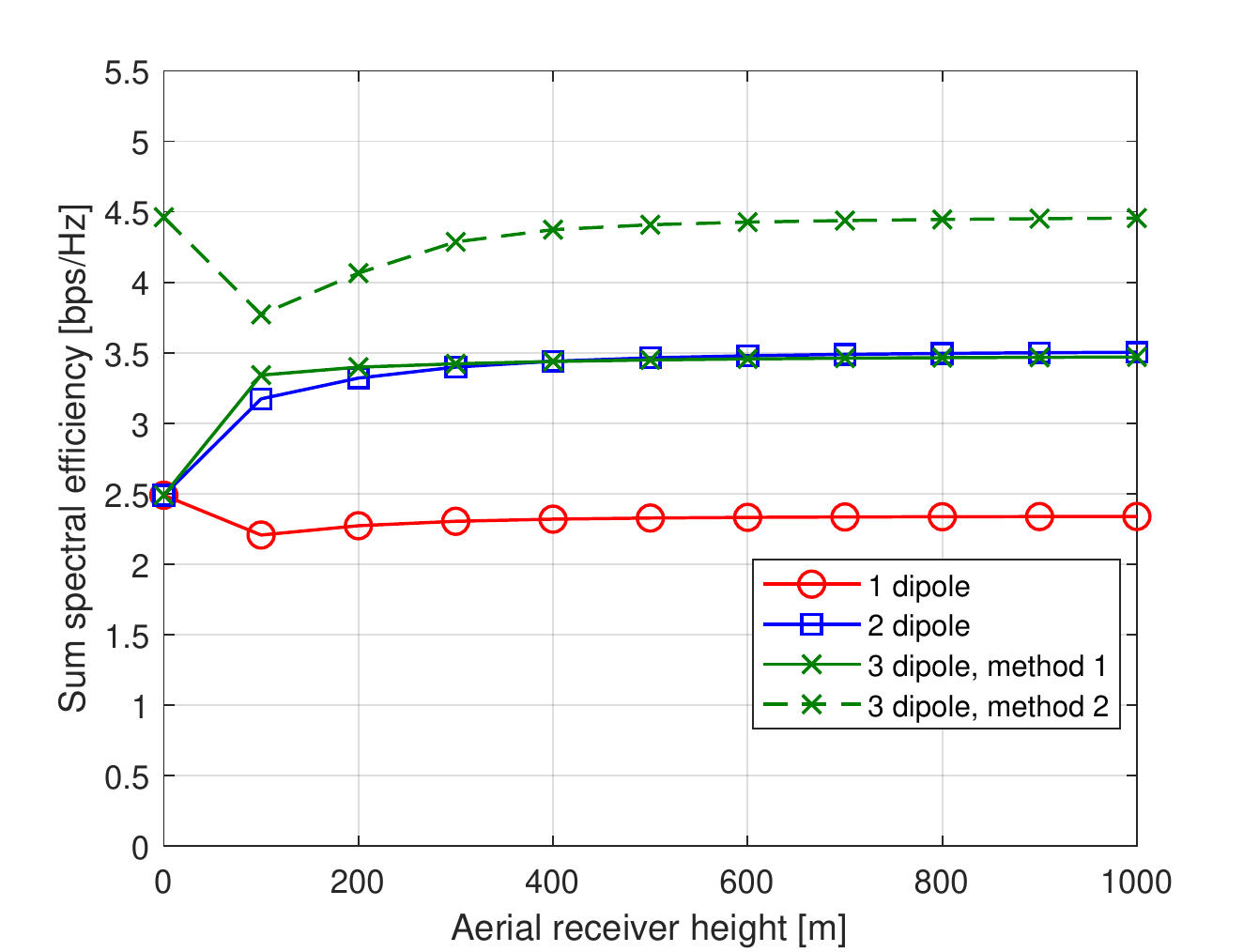}}
        \caption{Sum spectral efficiency vs air device height ($K = 4$).}
			\label{effi2}
\end{figure}
 \begin{figure}[t]
     \centering
     	\subfloat[A realization of 3D topology for IoT devices.]{
			\includegraphics[width=0.95\linewidth]{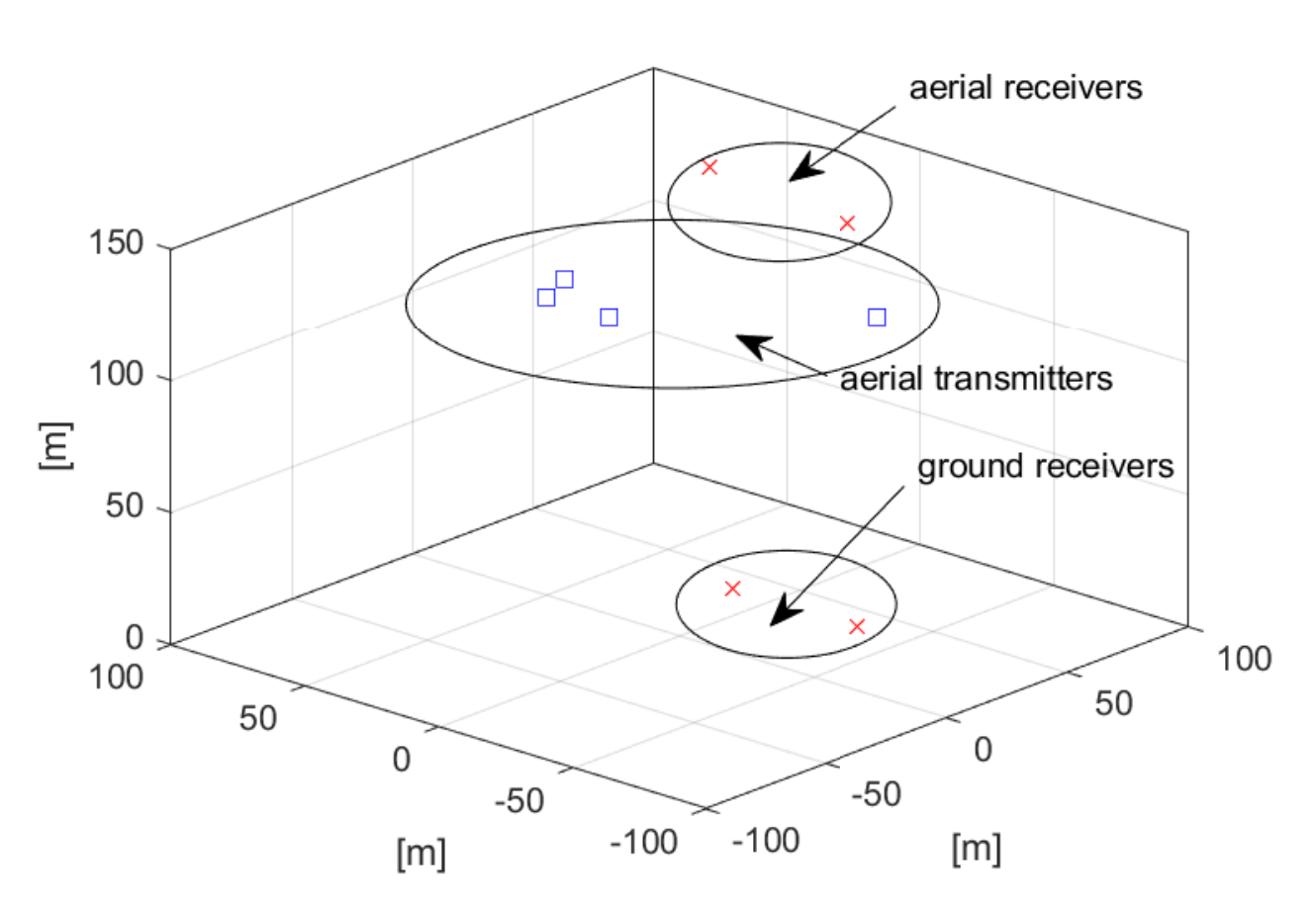}}
			
		\subfloat[Sum spectral efficiency.]{	\includegraphics[width=0.95\linewidth]{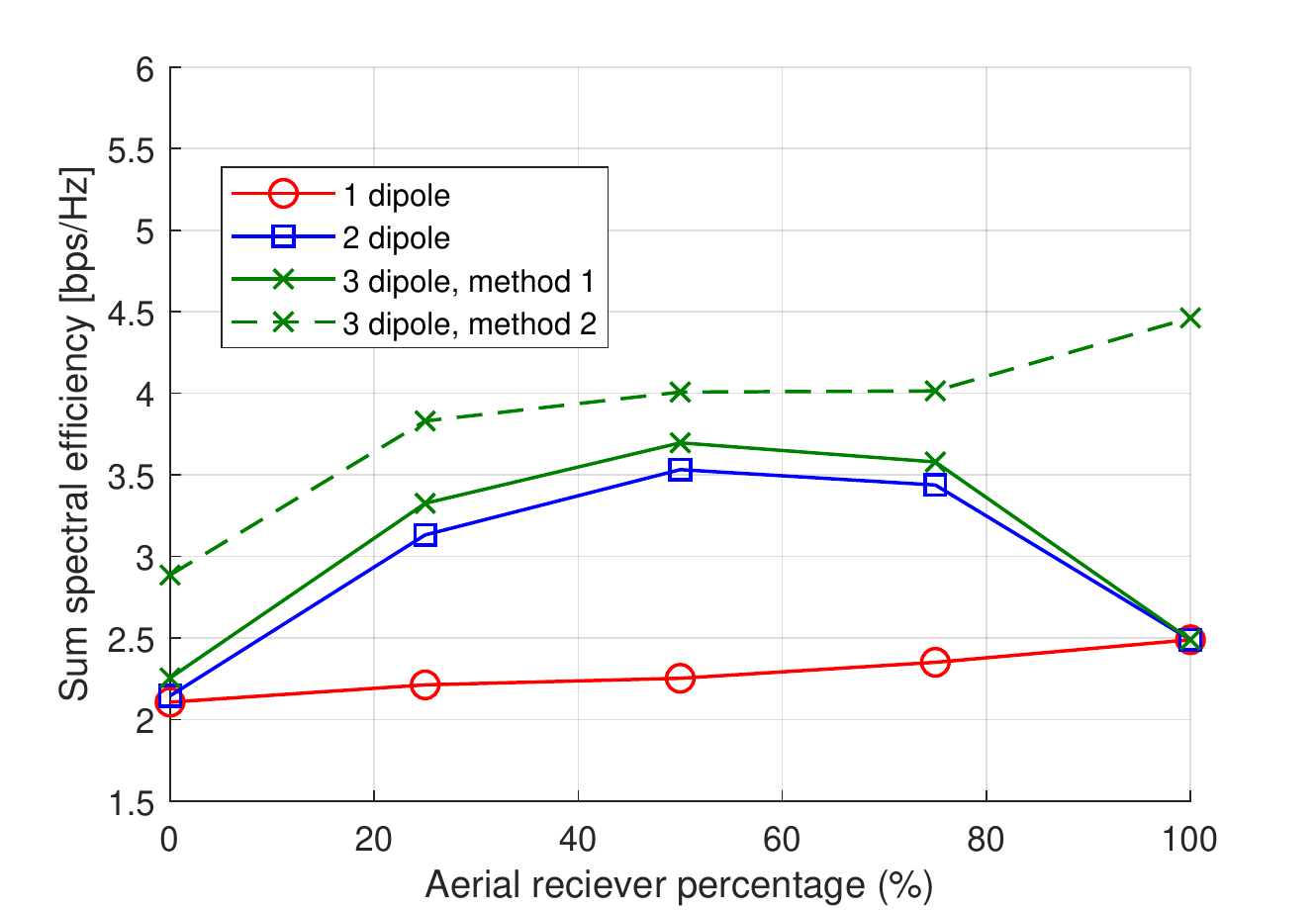}}
        \caption{Sum spectral efficiency of  aerial transmitter to aerial receiver scenario ($K = 4$, aerial transmitter percentage 100\%).}
			\label{effi3}
\end{figure}
 \begin{figure}[t]
     \centering
     	\subfloat[A realization of 3D topology for IoT devices.]{
			\includegraphics[width=0.95\linewidth]{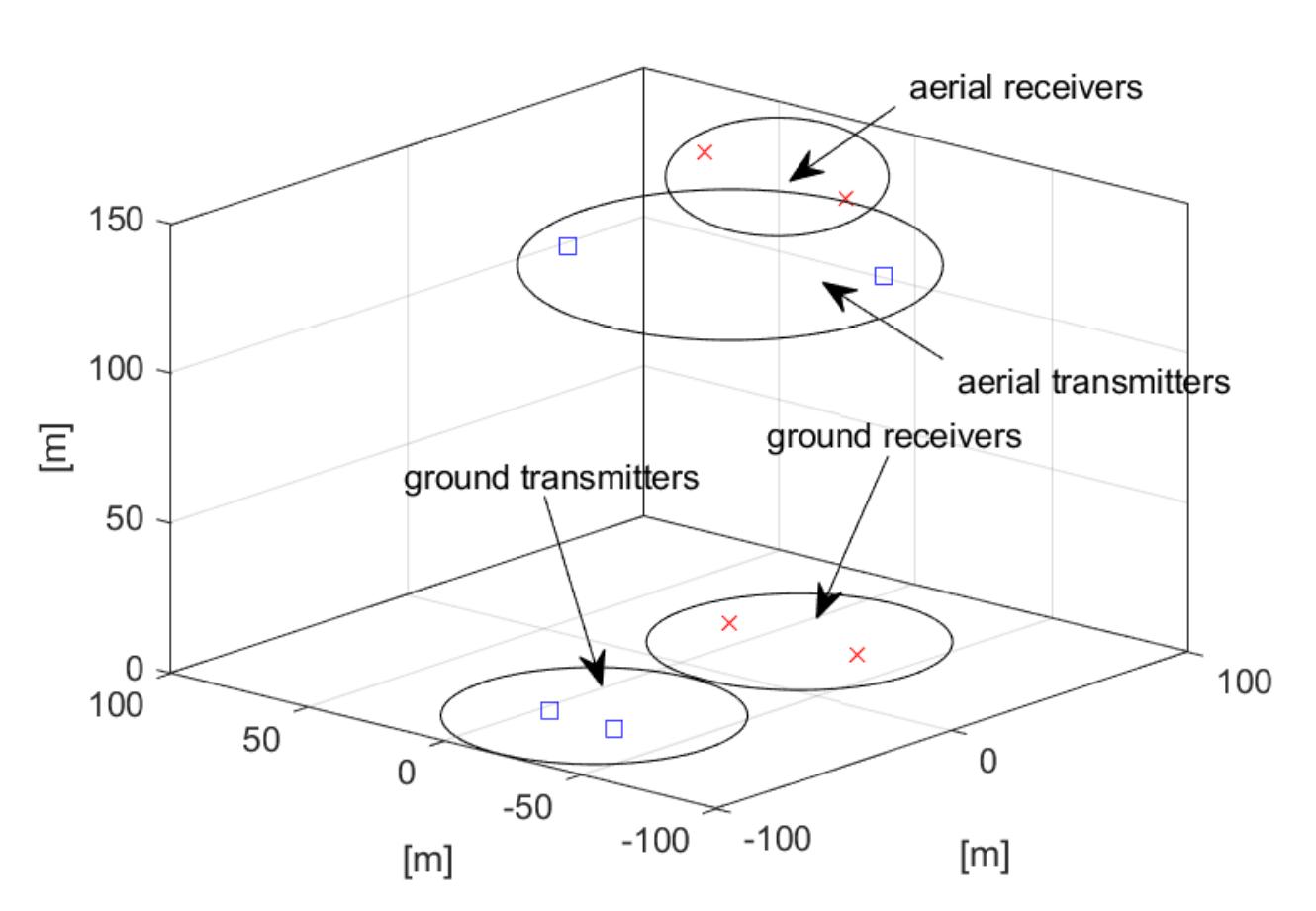}}
			
		\subfloat[Sum spectral efficiency.]{	\includegraphics[width=0.95\linewidth]{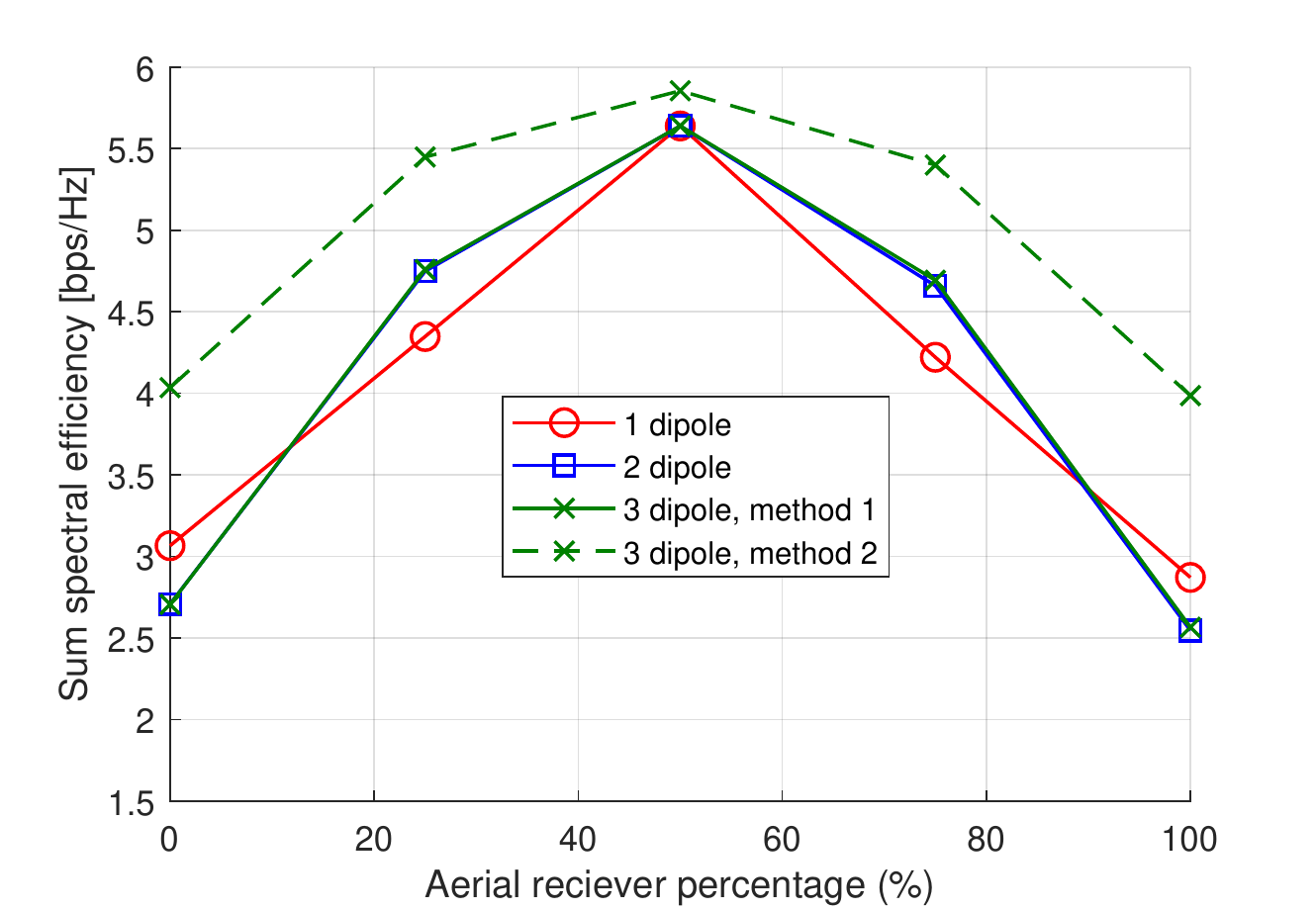}}
        \caption{Sum spectral efficiency of  aerial transmitter to aerial receiver scenario ($K = 4$, aerial transmitter percentage 50\%).}
			\label{effi4}
\end{figure}
From Fig.~\ref{effi}, we make the following observations:
\begin{itemize}
  \item The red curve shows the performance of single dipole antenna scenario. Since the radiation pattern is focused on ground devices, aerial devices have degradation in sum achievable rate. This tendency holds until 100 percentage of aerial receivers.
  \item The blue curve shows the performance of two cross dipole antenna scenario. The radiation pattern of y-axis dipole antenna is directed toward air devices, which enable transmitters to select better radiation pattern toward air devices. Compared with 1 dipole red curve, 2 dipole scenario achieves higher performance.
  \item In addition, the sum achievable rate of blue curve increases until around 50\% of aerial receivers. After that the performance decreases until 100\%  aerial receiver percentage. We can understand this observation from the fact that the directional radiation pattern can be utilized the best when the devices are well distributed in ground and air. In other words, if all aerial receivers are located at the either ground or air, we obtain bad performance.
  \item The green solid line shows the performance of three dipole scenario. In this case, we can utilize additional radiation patterns in order to increase performance.
  \item The green dotted line shows the performance of SLNR based antenna selection, which achieves the highest sum achievable rate. Since this approach selects optimal radiation pattern considering interference power as well as desired signal power, the performance is better than the green solid line curve.
  \end{itemize}
  \par
 In Fig.~\ref{pw}, we show the change of desired signal power, interference power, and the signal-to-interference-ratio as the height of aerial receivers increases. It is observed that both desired signal power and interference power decreases as the height of air device increases. There are two reasons for this observation. At first, path loss increases due to longer link distance from the height. Second, antenna gain is changed by different altitude. Since antenna gain keep decreasing in single dipole scenario, power degradation from height is huge, while since transmitters select the best antenna pattern in 2, 3 dipole scenario, power attenuation is weaker than dipole 1 scenario.\par
  Fig.~\ref{effi2} depicts sum capacity depending on the height of aerial receivers with different dipole configurations. We observe that the performance doesn't change much as the height increases in single dipole scenario, while higher sum capacity can be achieved by increasing the height in 2, 3 dipole scenario. This is due to the fact that gain from selecting the best antenna radiation pattern becomes higher as the height increases. Note that y-axis dipole radiation pattern shows higher gain in low elevation angle.\par
  In Fig.~\ref{effi3},~\ref{effi4}, we consider the scenario that aerial transmitters are communicating with aerial receivers. All transmitters are up in the air in Fig.~\ref{effi3}(a). The performance gap between '1 dipole' and '2 or 3 dipole' is the highest in the 50\% of aerial receivers in Fig.~\ref{effi3}(b), which shows a similar tendency with all ground transmitters scenario. In Fig.~\ref{effi4}, ground / air transmitters are equally distributed. In this setting, the '1 dipole' also achieve similar performance with '2 or 3 dipole' case, which utilizes additional directional antenna radiation patterns to improve performance in Fig.~\ref{effi4}(b).

\section{Conclusion}
In this work, we consider uncoordinated IoT network in 3D topology environment. We propose interference mitigation schemes with 3D radiation pattern of dipole antennas. The proposed schemes utilize the directivity of antenna radiation pattern when the number of dipole antennas is 2 or 3. In addition, we propose antenna selection methods in order to decide the best antenna pattern at the transmitter side. Simulation results verify the merit of the proposed schemes in both ground-to-air and air-to-air scenarios when IoT devices are well-distributed in the ground and air, and when the altitude of aerial devices is high.

\bibliographystyle{IEEEtran} 
\bibliography{ref}

\end{document}